# VISUAL, OPTICAL AND REPLICA INSPECTIONS: SURFACE PREPARATION OF 650 MHz NB CAVITY FOR PIP-II LINAC*

V. Chouhan†, D. Bice, D. Burk, M. K. Ng, G. Wu, Fermi National Accelerator Laboratory, Batavia, USA

## Abstract

Surface preparation of niobium superconducting RF cavities is a critical step for achieving good RF performance under the superconducting state. Surface defect, roughness, and contamination affect the accelerating gradient and quality factor of the cavities. We report surface inspection methods used to control the surface processing of 650 MHz cavities designated for the pre-production and prototype cryomodules for PIP-II linac. The cavity surface was routinely inspected visually, with an optical camera, and by microscopic scanning of surface replicas. This article covers details on the surface inspection methods and surface polishing process used to repair the surface.

## INTRODUCTION

Proton Improvement Plan II (PIP-II) program is an essential upgrade to the Fermilab's accelerator complex. This includes a superconducting RF (SRF) linear accelerator (linac) to accelerate a proton beam to 800 MeV. The PIP-II linac includes normal and superconducting sections. The superconducting section will incorporate different types of cavities including two types of 650 MHz 5-cell elliptical cavities called low-$\beta$ (0.61) and high-$\beta$ (0.92) 650 MHz cavities. The cavity surface is processed with the state-of-the-art techniques including electropolishing (EP), heat treatment at 800 or 900 °C, and final light EP for the baseline test performed in a vertical cryostat.

Surface defects have a detrimental effect on the performance of the cavities. The common defects found earlier were deep pits appeared as cat-eye feature, non-conformality of weld zone, rough surface etc. Optical inspection is typically employed to detect these defects. Previous studies have involved surface study of 1.3 GHz cavities using an optical camera developed at Kyoto University [1], which has proven effective in identifying surface features and defects [2]. To capture surface features, surface replicas are created using a molding process using a silicone liquid gel [2]. These replicas enable high-resolution microscopic study of small surface features present on the cavity surface [2, 3].

This manuscript focuses on the challenges encountered during the surface processing steps involved in the preparation of low-β (LB650) and high-β (HB650) elliptical cavities. Furthermore, it presents the crucial roles played by surface inspection techniques, such as visual inspection, optical inspection, and surface replica examination, in aiding the decision-making process for surface treatment.

## VISUAL AND CAMERA INSPECTIONS

Visual inspection of as-received cavity surface was performed to find any damage on the cavity surface. The visual inspection is performed to check the flanges, beam tube and the visible cell-wall through the beam tubes. Since the inner walls of the cells were not in line-of-sight through the beam tube, a small digital camera was inserted manually into the cavity to capture the images of the cell walls. Figure 1 shows a schematic of a digital camera inside the cavity. Each photograph covered a part of the half-cell. Multiple images were taken by rotating the camera along the cavity axis at every 45° angle. These images covered entire surface of the half-cell. The field of view of the camera covered the surface from the equator to iris of the half-cell. Photographs of another half-cell were captured when the camera angle was changed by 90° to face another half-cell. Figure 2 shows one of the captured photographs of the HB650 cavity before the EP process, providing sufficient clarity to detect any deep scratches, dents, or mechanical grinding marks. Except some minor scratches and mechanical polishing marks, no large defect was seen on the cavity surface. All the cavities qualified for the next step that was a detailed inspection of the equator surface with the optical camera.

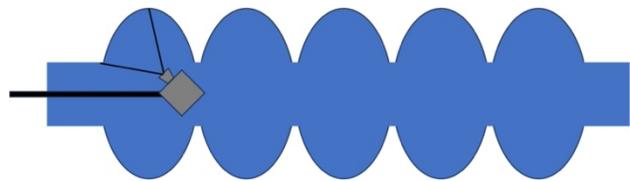

Figure 1: Schematic of a camera positioned inside the cavity.

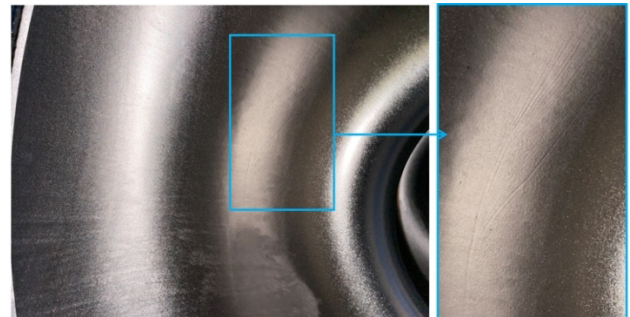

Figure 2: A part of the half-cell of an as-received HB650 cavity: Photograph captured with a digital camera inserted manually inside the cavity (left). Zoom-in image showing minor scratches (right).

_________________
* This work was supported by the United States Department of Energy, Offices of High Energy Physics and Basic Energy Sciences under contract No. DE-AC02-07CH11359 with Fermi Research Alliance.
† vchouhan@fnal.gov

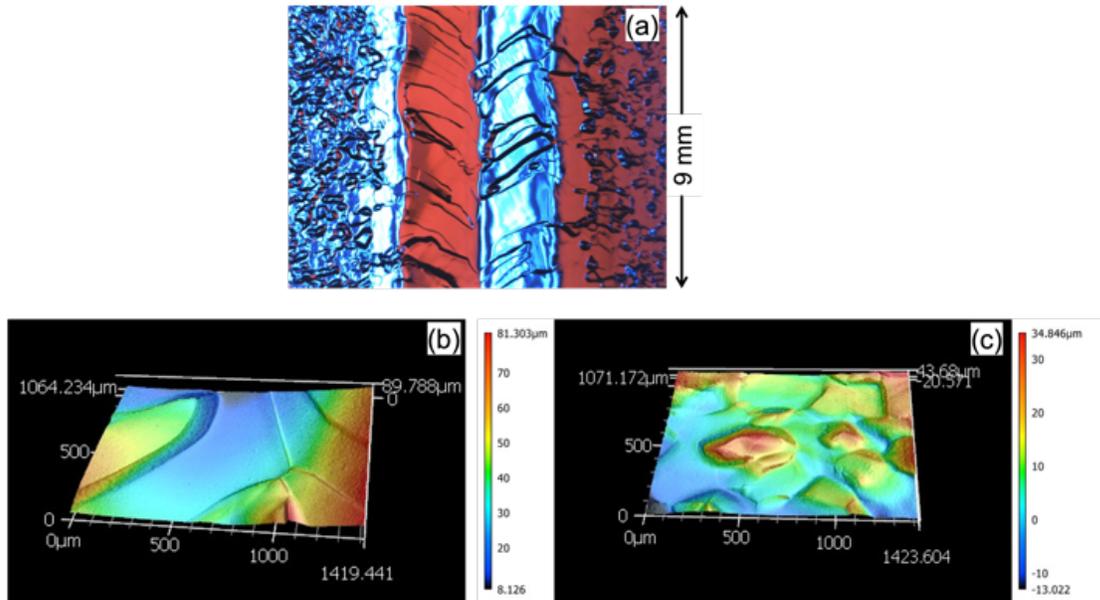

Figure 3: Optical image of the equator surface (a). LCSM images of replica surface representing the weld (b) and near-weld (c) equator positions.

## OPTICAL INSPECTION AND SURFACE REPLICA

The entire equator surface in all the five cells of as-received cavities was inspected with the optical camera. The optical camera provided an image having a dimension of 12×9 mm. The inspection aimed to detect any irregularities in the weld zone of equator and to find a defect that would make the cavity unqualified for undergoing any further surface processing. Although some features in the weld-zone were observed, all as-received cavities were considered to be processed.

A few cases are described in the following sections to report on the surface processing, challenges to prepare a smooth base surface, and stages at which surface inspection by optical camera and making replica were necessary.

### Surface Processing of LB650

Four LB650 cavities, namely B61C-EZ-101, -102, -103, and -104, that will be used in pre-production cryomodule were processed.

*B61C-EZ-103*: The cavity was electropolished with standard EP conditions [4,5]. The cavity quenched prematurely at 15 MV/m in a vertical test conducted at 2 K. Typically, such a low-field quench is associated with a significantly large defect on the equator surface.

To examine the surface, the entire equator surface in all 5-cells was scanned with the optical camera. A typical image of the equator surface is shown in Fig. 3. The optical inspection revealed a rough surface that differed from a standard electropolished surface. Although the optical images clearly showed that the surface was rough, it could not provide any information on roughness value. To obtain this information, a replica of the equator surface was created and examined using a laser confocal scanning microscope (LCSM). The replica images for weld and near-weld zones on equator are presented in Fig. 3 (b) and (c). Average roughness Ra of near-weld surface was estimated to be ~8 μm and grain step height was measured to be ~32 μm. The roughness was significantly higher than the buffered chemical polishing (BCP) surface which is usually found very rough with grain step height of ~5 μm. The rough surface might be responsible for the low quench field.

The information obtained with optical and replica images played a crucial role in evaluating the EP performance. It became evident that modifications to the electropolishing conditions were necessary to achieve a smoother surface. A study was conducted to optimize the EP parameters, details of which can be found elsewhere [4, 5]. The cavity after reset EP under optimized conditions met the $E_{acc}$ specification ($E_{acc}$ = 22.4 MV/m) set for the baseline test as reported elsewhere [5, 6].

*B61C-EZ-102*: This cavity was used for EP study to optimize the parameters. Before the study was started, the cavity already received unoptimized EP for 70 μm removal. During the parameters study, unoptimized conditions further removed 23 μm. Additional EP for 27 μm was applied with improved EP parameters.

Throughout the process of parameter optimization, optical inspections were performed at various stages to evaluate the surface yielded with different EP conditions. Figure 4 shows two images of the equator region, one after 70 μm removal under unoptimized conditions, and after 120 μm removal that included last 27 μm removal using the improved EP conditions. These images clearly demonstrate that the optimized EP process led to a smoother surface.

Based on this feedback, two new cavities (B61C-EZ-101 and -104) were processed using the optimized EP conditions, taking into account the surface smoothening observed in the images.

The cavity after surface processing completed with optimized EP satisfied the $E_{acc}$ specification for baseline test as reported in ref. [5, 6].

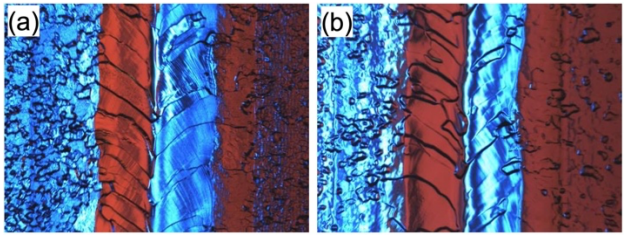

Figure 4: Optical image after unoptimized EP performed for 70 μm removal (b) Image after additional 23 μm removal with unoptimized EP and 27 μm removal with improved EP. Image size: 12×9 mm.

*B61C-EZ-101 and -104*: These cavities experienced optimum EP conditions in bulk and light EP [4, 5]. The cavity surface was examined with optical camera and by making replica. The replica that provided high-resolution features at the sub-micron level was inspected with LCSM. The optical and replica surface images after 120 μm removal are shown in Fig. 5. Average roughness Ra measured using LCSM was ~0.6 μm. The effectiveness of the optimized EP conditions was confirmed through these optical and replica inspections.

Both cavities after bulk removal of 120 μm were underwent a heat treatment at 800 °C and 40 μm light EP [4]. The cavity B61C-EZ-104 qualified in the baseline vertical test [5, 6]. Despite applying the optimized EP conditions, cavity B61C-EZ-101 quenched at 20 MV/m, failing to meet the specified baseline performance requirement in vertical tests. This early quenching may have been caused by a surface defect.

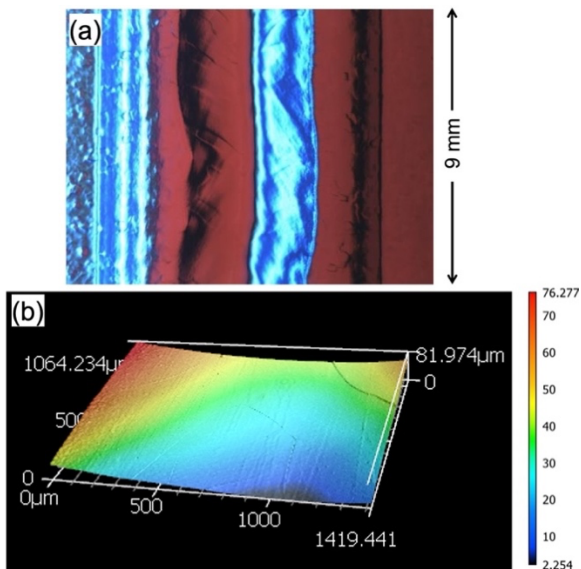

Figure 5: Optical image of typical equator surface of B61C-EZ-104 (a) and LCSM image of the replica representing the equator surface of B61C-EZ-101.

The entire equator surface in all five cells was inspected with the optical camera. A unique feature on the equator surface in cell-1 was found as shown in Fig. 6 (a). The feature found in the optical inspection was unclear to decide the next surface processing.

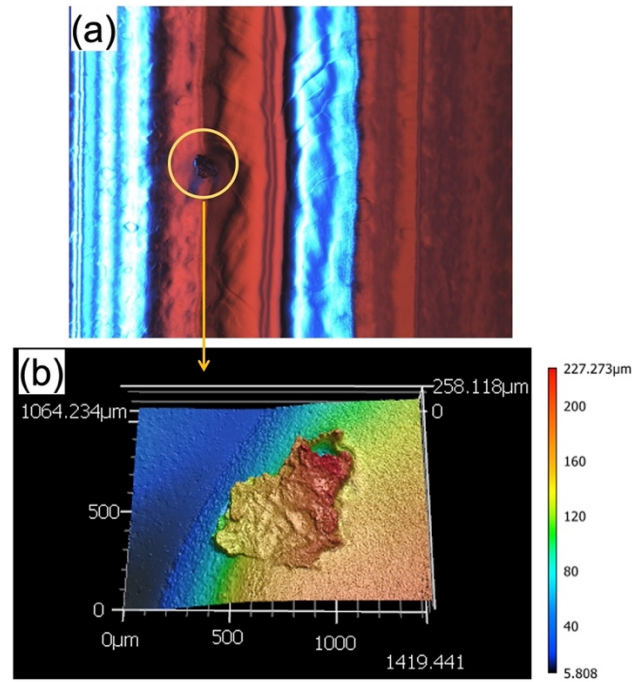

Figure 6: (a) Optical image showing a unique feature (image size is 12×9 mm). (b) LCSM image of replica representing the surface in (a).

To get a detailed information of the feature, replica of the surface was made. An LCSM image of the feature captured by replica is shown in Fig. 6 (b). The image represented that the feature on the cavity surface was a random-shaped crater with a depth of ~170 μm. The deep crater with sharp edges was identified as the likely cause of the early quenching.

Given the size and depth of the feature, employing additional electropolishing (EP) was deemed unsuitable for removing the surface defect. Therefore, the decision was made to do local mechanical grinding to remove the defect. The subsequent sections will provide further details on the specific approach taken for grinding.

*Local grinding and post-griding process (B61C-EZ-101):* The defect spot in the end cell of cavity B61C-EZ-101 was manually polished with aluminum oxide ($Al_2O_3$) abrasive media while the cavity was sitting on the optical stand for surface inspection during the different polishing steps performed with different grit size papers. The polishing paper was fixed on a handy tool having a rubber pad. This rubber pad holding the polishing paper might minimize embedding of particles into the Nb surface. The final polishing step utilized a grit size of 320. An optical image of the polished surface to compare with the original image (Fig. 6) is shown in Fig. 7. The image revealed that the defect was successfully polished, resulting in a smooth surface with traces of abrasive particles. Optical inspections conducted at various stages of the polishing process served as guidance to determine when to stop the polishing procedure.

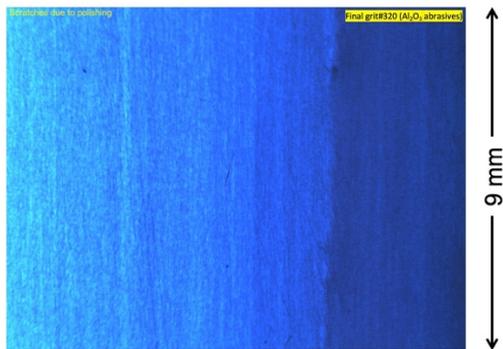

Figure 7: Optical image of the defect position shown in Fig. 6 after local grinding with Al$_2$O$_3$ abrasive media.

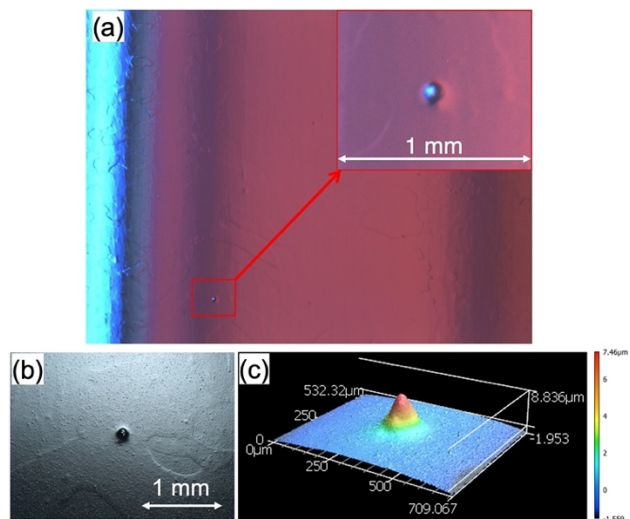

Figure 8: (a) Optical image of equator surface mechanically polished followed by EP for 60 μm removal. (b) Optical image of replica representing surface in (a). (c) 3D image of the negative replica showing the pit.

The next processing step was chosen to be electropolishing for a removal thickness of 90 μm. This removal depth was considered sufficient to remove embedded particles out from the Nb surface. Following the EP process, an optical inspection was conducted to assess the presence of any remaining defects on the surface. The inspection revealed the presence of certain polishing features, including pit- or bead-like structures in the polishing region, as shown in Fig. 8 (a).

Given that optical inspection alone could not definitively determine whether these features were pits or embedded particles, a surface replica was made to confirm the same. The replica image apparently showed that the cavity surface had a pit having a diameter of ~200 μm and a depth of ~7 μm. It is likely that this pit formed when a large-sized embedded particle detached from the surface during the EP process. The results suggested that the polishing media and the grit size should be properly selected to avoid such pit formation on the surface. Based on the depth of the observed pits, 20 μm removal in EP was decided, where each 10 μm was removed before and after subjecting the cavity to a heat treatment at 900 °C for 3 hours.

The cavity met the $E_{acc}$ specification desired in the baseline performance and was ready for high-Q treatment. The test result is reported in [6].

### Surface Processing of HB650

*B92F-RI-203*: Optical inspection of the cavity surface was performed after 160 μm removal in EP conducted with the conditions specified in reference [7]. A defect appeared as a cat-eye feature on equator surface in cell-1 of the cavity. The cat-eye feature is usually treated by local mechanical grinding or centrifugal barrel polishing. To obtain detail of the cat-eye feature, replica of the surface was prepared. The optical image of the surface and the corresponding replica image are shown in Fig. 9 (a) and (b). The feature was a wide pit with depth of ~25 μm, respectively. This pit indicated the presence of an underlying defect in the Nb bulk, which became apparent as a pit following the 160 μm removal during the EP process.

Since the cat-eye feature had no sharp edges, the surface was not mechanically polished. The cavity was processed with 900 °C followed by EP for 10 μm removal.

The cavity reached 29.3 MV/m at $Q_0$ of 2.2×10$^{10}$ and met the specification ($E_{acc}$ = 25 MV/m) set for HB650 cavities in their baseline tests. The detail of EP and $Q_0$ versus $E_{acc}$ curve is shown in reference [7].

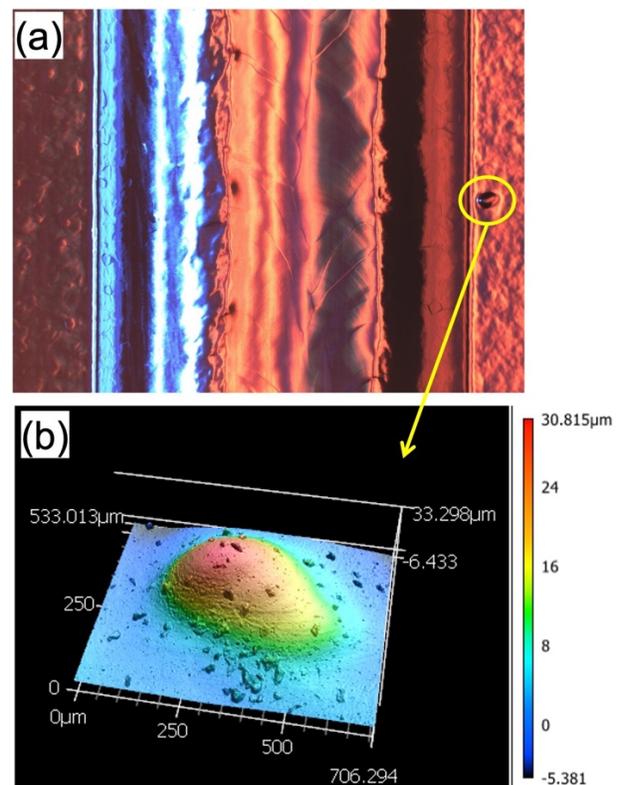

Figure 9: (a) Optical image showing a cat-eye feature on equator surface in cell-1 of B92F-RI-203. (b) 3D image of the cat-eye feature captured with surface replica.

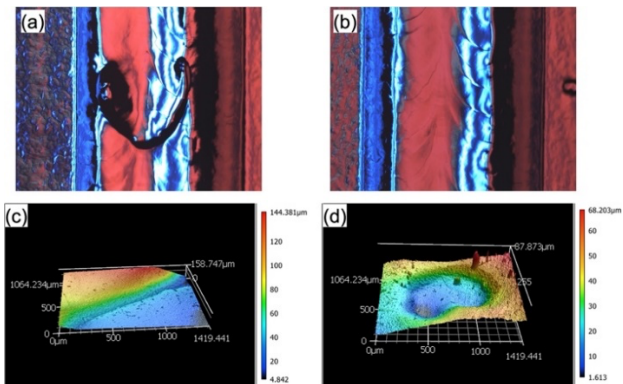

Figure 10: Optical images (12×9 mm) (a) showing defect on weld-zone and (b) a defect near weld-zone of equator. (c) Replica image of the defect in (a). (d) Replica image of the defect in (b).

*B92F-RI-202*: The equator surface, as observed with optical camera, of the as-received cavity showed some weld features on the equator in all the five cells. An optical inspection performed after 167 µm removal in the EP process showed that the features remained on the surface as shown in Fig. 10 (a). An additional defect on the equator surface in cell-5 was found (Fig. 10 (b)). The replica of both features was produced. The LCSM images of the replica surface is shown in Fig. 10 (c) and (d).

The replica image revealed that the edges of the feature were not sharp. Considering the feature details in the replica, the cavity surface was processed to prepare for a baseline test. The vertical test result will be reported later.

*B92D-RRCAT-502*: The cavity was processed with EP for 150 µm removal, N-doping, and post-doping EP for 5 µm removal. The cavity tested after the processing was prematurely quenched at 11.2 MV/m. In optical inspection, the cavity surface in cell-5 showed a pit that was polished by local grinding followed by 60 µm EP, heat treatment at 800 °C for 3 h and light EP for 10 µm. The pit was still present on the surface as found in an optical inspection (Fig. 11).

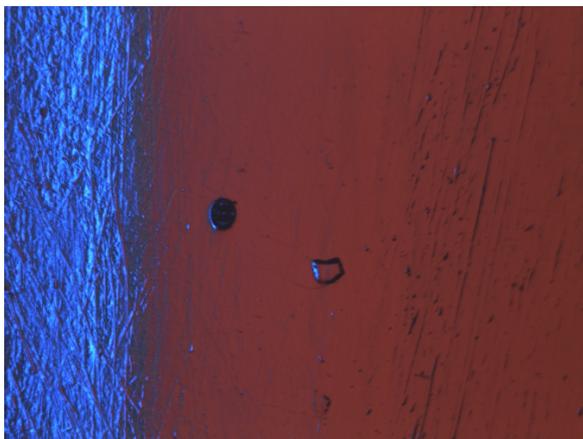

Figure 11: Defect on the equator surface in cell-5 of B92D-RRCAT-502. Image size: 12×9 mm.

Replica was made to get the pit feature. The diameter and depth of the pit were ~450 µm and 160 µm, respectively. Similar to the grinding process applied to B61C-EZ-101, local grinding on the defect position was performed with $Al_2O_3$ abrasive media having grit size of 80–400 until the defect disappeared from the surface. Following the grinding, EP for 60 µm was applied to the cavity. The cavity reached at 29.8 MV/m and met the baseline specification.

## CONCLUSION

We reported the challenges faced in preparation of the LB650 and HB650 5-cell cavities for the PIP-II linac. To decide an appropriate processing step to the cavities specially after being disqualified due to not meeting performance specifications, surface inspection becomes crucial. The surface and replica inspections were performed at different stages of the EP optimization process. The surface inspection process including visual, optical, and replica were used to attain the surface information before and after the surface processing as well as to decide an appropriate surface process step depending on the surface status. The optical and replica inspections were instrumental in optimizing the EP parameters and mechanical polishing process.

## ACKNOWLEDGEMENTS

This manuscript has been authored by Fermi Research Alliance, LLC under Contract No. DE-AC02-07CH11359 with the U.S. Department of Energy, Office of Science, Office of High Energy Physics.

Grand Rapids, MI, USA, Jun. 2023, paper TUPTB042, this conference.